\documentclass[prb,twocolumn,showpacs,amsmath,amssymb,superscriptaddress,floatfix]{revtex4}
\pdfoutput=1

\usepackage{ graphicx } 
\usepackage{ bm } 
\usepackage{ color }

\newcommand{\ua}{\uparrow}
\newcommand{\da}{\downarrow}

\newcommand{\be}{\begin{equation}}
\newcommand{\ee}{\end{equation}}
\newcommand{\bea}{\begin{eqnarray}}
\newcommand{\eea}{\end{eqnarray}}

\newcommand{\bs}{\boldsymbol}

\newcommand{\tr}{\mbox{Tr}}

\newcommand{\vk}{{\boldsymbol k}}

\begin{document}
\title{Vanishing $k$-space fidelity and phase diagram's bulk-edge-bulk correspondence}
\author{ P. D. Sacramento$^1$, B. Mera$^{1,2}$ and N. Paunkovi\'c}
\affiliation{\textit CeFEMA,
Instituto Superior T\'ecnico, Universidade de Lisboa, Av. Rovisco Pais, 1049-001 Lisboa, Portugal }
\affiliation{Instituto de Telecomunica\c{c}\~oes, 1049-001 Lisboa, Portugal}
\affiliation{Departmento de Matem\'atica, Instituto Superior T\'ecnico, Universidade de Lisboa, Av. Rovisco Pais, 1049-001 Lisboa, Portugal}

\date{ \today }


\begin{abstract}
The fidelity between two infinitesimally close states or the fidelity susceptibility of a system are known to detect quantum phase transitions. Here we show that the $k$-space fidelity  between two states far from each other and taken deep inside (bulk) of two phases, generically vanishes at the k-points where there are gapless points in the energy spectrum that give origin to the lines (edges) separating the phases in the phase diagram. We consider a general case of two-band models and present a sufficient condition for the existence of gapless points, given there are pairs of parameter points for which the fidelity between the corresponding states is zero. By presenting an explicit counter-example, we showed that the sufficient condition is not necessary. Further, we showed that, unless the set of parameter points is  suitably constrained, the existence of gapless points generically imply the accompanied pairs of parameter points with vanishing fidelity. Also, we showed the connection between the vanishing fidelity and gapless points on a number of concrete 
examples (topological triplet superconductor, topological insulator, $1d$ Kitaev model of 
spinless fermions, BCS superconductor, Ising model in a transverse field, graphene and Haldane Chern insulator), 
as well as for the more general case of Dirac-like Hamiltonians. We also briefly discuss the 
relation between the vanishing fidelity and gapless points at finite temperatures. 

\end{abstract}


\maketitle

\section{Introduction}

The fidelity and other quantum information signatures have been used to
distinguish and characterize quantum phases, with particular emphasis
in signalling their transitions \cite{zanardi,gu}. Traditionally one compares states
that differ infinitesimally due to some change of parameters of the
Hamiltonian or due to some change in temperature or other intensive
quantities associated with some reservoirs. The results together with
some generalization, such as partial state fidelity \cite{paunkovic1,zhou}
or the fidelity spectrum \cite{sacramento1,gu2}, have been 
used to detect quantum phase transitions including those of a topological nature.
This includes topological insulators and topological superconductors 
\cite{kane,zhang,alicea,schnyder}.

\begin{figure}
\includegraphics[width=0.75\columnwidth]{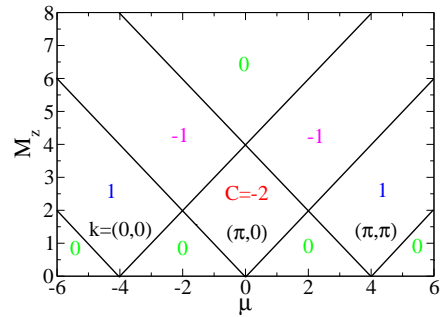}
\caption{\label{fig1}
(Color online) Phase diagram of a triplet superconductor as a function of chemical
potential and Zeeman term. $C$ is the Chern number. $k$ is the momentum of each
transition line.
}
\end{figure}

The procedure was used to study the topological phases
and transitions in various systems 
\cite{min,chen77,gu77,hamma77,abasto78,yang78,campus78,abasto79,zhao80,eriksson,castelnovo,trebst,wang10,wcyu,eisertrmp,balents,kitaev96,levin96,balents2,misguich,depenbrock,kallin,Yao} 
and, in particular, in a two-dimensional triplet superconductor \cite{sato}, which displays
several trivial and topological phases, labelled by Chern numbers
or a $Z_2$ invariant. Spinfull electrons in the presence of a Zeeman term
(that breaks time reversal symmetry) and in the presence of Rashba spin-orbit
coupling are in a superconducting state with both singlet and triplet pairing
symmetry (parity is broken due to the presence of the spin-orbit coupling).
The Hamiltonian is written as
\begin{eqnarray}
\hat H = \frac 1 2\sum_\vk  \left( {\boldsymbol c}_{\vk}^\dagger ,{\boldsymbol c}_{-\vk}   \right)
\left(\begin{array}{cc}
\hat H_0(\vk) & \hat \Delta(\vk) \\
\hat \Delta^{\dagger}(\vk) & -\hat H_0^T(-\vk) \end{array}\right)
\left( \begin{array}{c}
 {\boldsymbol c}_{\vk} \\  {\boldsymbol c}_{-\vk}^\dagger  \end{array}
\right)
\label{bdg1}
\end{eqnarray}
where $\left( {\boldsymbol c}_{\vk}^{\dagger}, {\boldsymbol c}_{-\vk} \right) =
\left( c_{\vk\ua}^{\dagger}, c_{\vk\da}^\dagger ,c_{-\vk\ua}, c_{-\vk\da}   \right)$
and
\bea
\hat H_0 &=& \epsilon_\vk\sigma_0 -M_z\sigma_z + \hat H_R\,.
\nonumber \\
\hat H_R &=& \boldsymbol{s} \cdot \boldsymbol{\sigma} = \alpha
\left( \sin k_y \sigma_x - \sin k_x \sigma_y \right)\,,
\eea
Here, $\epsilon_{\boldsymbol{k}}=-2 t (\cos k_x + \cos k_y )-\mu$
is the kinetic part, $t$ denotes the hopping parameter set in
the following as the energy scale, $\mu$ is the
chemical potential,
$\boldsymbol{k}$ is a wave vector in the $xy$ plane, and we have taken
the lattice constant to be unity. $M_z$
is the Zeeman splitting term responsible for the magnetization,
in energy units and the $\hat H_R$ is the Rashba spin-orbit term. 
$\alpha$ is measured in the energy units
and $\boldsymbol{s} =\alpha(\sin k_y,-\sin k_x, 0)$.
The matrices $\sigma_x,\sigma_y,\sigma_z$ are
the Pauli matrices acting on the spin sector, and $\sigma_0$ is the
$2\times2$ identity.
The pairing matrix reads
\begin{equation}
\hat \Delta = i\left( {\boldsymbol d}\cdot {\boldsymbol\sigma} + \Delta_s \right) \sigma_y =
 \left(\begin{array}{cc}
-d_x+i d_y & d_z + \Delta_s \\
d_z -\Delta_s & d_x +i d_y
\end{array}\right)\,.
\end{equation}
The system has a rich phase diagram with trivial and topological phases.
These are show in Fig. \ref{fig1} considering $d_z=0$ and choosing
$d_x=\Delta_t \sin k_y, d_y=-\Delta_t \sin k_x$.
The Hamiltonian studied has therefore in general a $4 \times 4$ matrix structure.
The problem is easilly diagonalized and the lines where gapless points occur
separate the different topological phases. 

This model was studied in Refs. \onlinecite{tharnier,tee}.
A particular interest
was the study of entanglement and fidelity. These quantities were determined
by numerical diagonalization of density matrices and the fidelity.
Since the model factorizes in $k$-space the fidelity may be calculated
for each momentum separately. In addition to the usual sensitivity of the fidelity
around the critical points it was noted that some signature of these critical
lines emerges at specific momentum values (associated with the points where
the gap vanishes and the transitions occur) such that the $k$-space fidelity
vanishes. This occurs even though it is calculated with density matrices that correspond
to points in the phase diagram that are deep inside the various phases and
not necessarily in the vicinity of the transition lines.
 
In this work we aim to understand better this result.

We begin by noting that the topology is not changed if $\Delta_s=0, \alpha=0$ as shown in \cite{sato}. If we take
these values the $4\times 4$ matrix decouples in two $2\times 2$ matrices since
the spin components do not get mixed anymore as
\begin{equation}
H_{\uparrow \uparrow} = 
 \left(\begin{array}{cc}
\epsilon_k-M_z & -i \Delta_t \left(\sin k_x-i \sin k_y \right) \\
i \Delta_t \left(\sin k_x+i \sin k_y \right) & -\epsilon_k+M_z
\end{array}\right)\,.
\end{equation}
and
\begin{equation}
H_{\downarrow \downarrow} = 
 \left(\begin{array}{cc}
\epsilon_k+M_z & -i \Delta_t \left(\sin k_x+i \sin k_y \right) \\
i \Delta_t \left(\sin k_x-i \sin k_y \right) & -\epsilon_k-M_z
\end{array}\right)\,.
\end{equation}

These matrices can be written in terms of Pauli matrices as
\bea
H_{\uparrow \uparrow} &=& (\epsilon_k-M_z) \sigma_z - \Delta_t \sin k_y \sigma_x
+\Delta_t \sin k_x \sigma_y \nonumber \\
H_{\downarrow \downarrow} &=& (\epsilon_k+M_z) \sigma_z + \Delta_t \sin k_y \sigma_x
+\Delta_t \sin k_x \sigma_y \nonumber \\
\eea
Denoting a vector $\bs{h}_{\sigma=\uparrow}=\bs{h}_{\uparrow \uparrow}$ and
$\bs{h}_{\sigma=\downarrow}=\bs{h}_{\downarrow \downarrow}$
we get that the Hamiltonian matrices may be written in the form
$\bs{h}_{\sigma} \cdot \bs{\sigma}$, with
\be
\bs{h}_{\uparrow}=(-\Delta_t \sin k_y,\Delta_t \sin k_x,\epsilon_k-M_z)
\label{hup}
\ee
and
\be
\bs{h}_{\downarrow}=(\Delta_t \sin k_y,\Delta_t \sin k_x,\epsilon_k+M_z).
\label{hdown}
\ee

The reduction of the problem to two $2\times 2$ matrices simplifies
the problem considerably and an analytical solution for the fidelity is
easy to obtain. Its analysis clarifies that the vanishing of the $k$-space
fidelity at selected points is a general feature associated to a gapless point.
We verify this result considering several models that display transitions either
topological or non-topological.

In section II we recall the definition of the fidelity and the
$k$-space fidelity both in the finite temperature and zero temperature
regimes. Then we apply it to the $2d$ triplet topological superconductor
emphasizing the connection between the momenta where the fidelity vanishes
and the spectrum gapless points at the transition lines.    
We perform an apstract analysis of the relation between zero-fidelity and gapless points, for the case of a general $2 \times 2$ Hamiltonian.
Other models are considered also at zero temperature, both topological and non topological. In section~\ref{sec:application}, models such as  topological insulators, $1d$ Kitaev model of spinless fermions and the Haldane Chern insulator are discussed. Further, in Appendix~\ref{sec:other_applications} we analyze a conventional superconductor, the Ising model in a transverse field and graphene. Also, in Appendix~\ref{sec:temperature} the triplet $2d$ superconductor is considered at finite temperature showing that as temperature decreases the $k$-space fidelity approaches the regime of vanishing points at the gapless points that occur at the transition lines between different phases. In section IV we consider a generalization to higher dimensional Hamiltonians. The zero temperature fidelity is obtained and applied to a $3d$ topological insulator, further establishing the connection between transition lines with gapless points and zeros in the $k$-space fidelity. 
The reverse however is not always true. It is possible to find models where, although vanishing points 
in the fidelity can correspond to gapless excitations, they are not associated with transition lines. 
This is shown in section V for a normal non-topological tight-binding model with a Zeeman term.  
The fidelity vanishes in extended regions that correspond to gapless points in the spectrum that 
are not associated with transition lines between phases. Another example is also considered that 
leads to a vanishing fidelity as a function of some control parameter introduced in a model of 
graphene that allows a continuous transition between the two opposite poles of the $\bs{h}$ 
vector Hamiltonian, with no specific gapless point in momentum space, since the spectrum 
vanishes for all momenta. We conclude with section VI.

\section{Fidelity}

The quantum
fidelity between two pure states (for two sets of parameters) is
the absolute value of the overlap between the ground states for the two sets of parameters.
In general, the quantum fidelity \cite{Jozsa} between two states characterized by two density
matrices $\rho_1$ and $\rho_2$, may be defined as the trace of the fidelity operator,
${\cal F}$, as
$F(\rho_1,\rho_2) = \text{Tr} {\cal F}= \text{Tr} \sqrt{ \sqrt{\rho_1} \rho_2 \sqrt{\rho_1}}$.
The fidelity operator $\mathcal{F}$ can be studied using different basis states, associated with
different representations, such as position, momentum, energy or charge and spin.

\begin{figure}
\includegraphics[width=0.75\columnwidth]{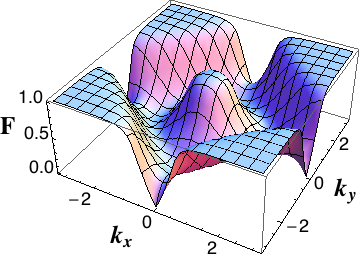}
\caption{\label{fig2}
(Color online) $k$-space fidelity for the $2d$ triplet superconductor with 
$\Delta_{t,1}=\Delta_{t,2}=0.6, \mu_1=-3,\mu_2=-0.1, M_{z,1}=M_{z,2}=0.5, T=0$. 
}
\end{figure}

\begin{figure}
\includegraphics[width=0.48\columnwidth]{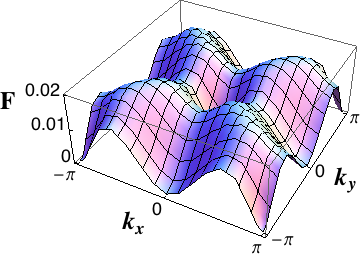}
\includegraphics[width=0.48\columnwidth]{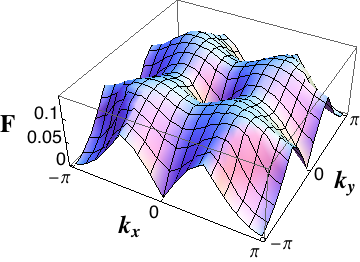}
\caption{\label{fig3}
(Color online) 
$k$-space fidelity for the $2d$ triplet superconductor with 
$\Delta_{t,1}=\Delta_{t,2}=0.6$, (left panel) $\mu_1=-6.0,\mu_2=6.0, M_{z,1}=M_{z,2}=0.5, T=0$. 
One gets the same result for $M_{z,1}=M_{z,2}=0$.
In the right panel $\mu_1=-2.0,\mu_2=6.0, M_{z,1}=4, M_{z,2}=0.5, T=0$. 
The behavior of the fidelity has linear dispersion around
$\bs{k}=(0,0),(\pi,0)$ and quadratic around $\bs{k}=(\pi,\pi)$.
}
\end{figure}

Since the Hamiltonian is separable in momentum space,
the density matrix operator for a momentum $k$ may be defined as usual as
\begin{equation}
\label{ }
        \hat{\rho}_{k}  = \frac { \mathrm{e}^{-\beta \widehat{H}_{k} }  } { Z_{k} },
\end{equation}
In the diagonal basis it is written as
\begin{equation}
\label{ }
        \rho_{k}=\left < n | \hat{\rho}_{k} | n \right > =
        \frac { \mathrm{e}^{-\beta \left< n | \widehat{H}_{k} | n \right >  }  } { Z_{k} } .
\end{equation}

In Ref. \onlinecite{tharnier}
a basis representation for the density matrix in terms of
the occupation numbers for a given momentum (and its symmetric) and the two spin
projections was used.
The eigenvalues of the density matrix are obtained if we diagonalize the Hamiltonian in the same basis.
We considered the representation
\begin{equation}
\label{eq:hamiltonian-new-base-state}
        \widetilde{H}_k =\left < n_{k_{\uparrow}}  n_{- k_{\uparrow}}  
   n_{k_{\downarrow}}  n_{-k_{\downarrow}} \right |
        \widehat{H}_{k} \left | n_{k_{\uparrow}}  n_{- k_{\uparrow}}  n_{k_{\downarrow}}  n_{-k_{\downarrow}} \right >
\end{equation}
The diagonalization of the Hamiltonian matrix in this enlarged basis is written as
\begin{equation}
\label{ }
        \widetilde{H}_k \bm{Q}_{k,n}=\lambda_{k,n} \bm{Q}_{k,n} \quad ; \quad n = 1, \ldots, 16
\end{equation}
note that $n$ here is just an index number and should not be confused with the occupation number of equation (\ref{eq:hamiltonian-new-base-state}).
In the same basis the density matrix may be written as
\begin{equation}
\label{ }
        \rho_{k}=\frac { \mathrm{e}^{-\beta \widetilde{H}_{k} }  } { Z_{k} } .
\end{equation}
Therefore the eigenvalues of the density matrix may be written as
$\rho_k \bm{Q}_{k,n} = \Lambda_{k,n} \bm{Q}_{k,n}$
where
\begin{equation}
\label{eq:eigval-DM}
        \Lambda_{k,n} = \frac { \mathrm{e}^{ -\beta \lambda_{k,n} } } { \sum\limits_{n'} \mathrm{e}^{ -\beta \lambda_{k,n'} } } .
\end{equation}

A simpler way to calculate the fidelity is to use a representation in the basis
of the creation and destruction operators. In the case of the Sato and Fujimoto
model this leads to a $4\times 4$ representation (the same dimension of the
Hamiltonian matrix).

As mentioned in the introduction the problem may be further simplified noting that
the density matrix for the Sato and Fujimoto model with $\Delta_s=\alpha=0$ may be written as
\bea
\rho &=& \rho_{\uparrow} \rho_{\downarrow} \nonumber \\
\rho_{\sigma} &=& \frac{\prod_k e^{-\beta H_k^{\sigma}}}{\prod_k 
\text{Tr} \left( e^{-\beta H_k^{\sigma}}\right)}
\eea
where now the matrices have a dimension $2\times 2$.


The case of an Hamiltonian with a $2\times 2$ structure has been considered
before \cite{paunkovic2,mera1,mera2}.
The $k$\space fidelity between two states $\rho_1$ and $\rho_2$ can be written as
\be
F_{12}(\bs{k}) = \frac{2+\sqrt{2\left(1+A_{12}(k)+
B_{12}(k) \bs{n}_{k,1} \cdot \bs{n}_{k,2} \right)}}{
\sqrt{(2+2\cosh (\beta E_{k,1}/2) ) (2+2 \cosh (\beta E_{k,2}/2))}}
\ee
where $E_{k,i}$ are the energy eigenvalues of the Hamiltonian $H_i$ and
\bea
A_{12}(k)&=&\cosh (\beta E_{k,1}/2) \cosh (\beta E_{k,2}/2) \nonumber \\
B_{12}(9)&=&\sinh (\beta E_{k,1}/2) \sinh (\beta E_{k,2}/2)
\eea

In the zero temperature limit $\beta \rightarrow \infty$ the expression simplifies.
The fidelity is given by
\be 
F_{12}(\bs{k}) =  
\sqrt{ \frac{1}{2} \left( 1+ \frac{\bs{h}_1}{|\bs{h}_1|} \cdot
\frac{\bs{h}_2}{|\bs{h}_2|} \right) } 
\label{central}
\ee
In this case $E_k=|\bs{h}|$.
It is easy to see that if $\bs{h}_1=\bs{h}_2$ the fidelity is one.

\subsection{k-space fidelity of $2d$ triplet superconductor}

We begin by considering the case of zero temperature.
Recalling that
\be
F_{12} = F_{12}^{\uparrow} F_{12}^{\downarrow}
\ee
the fidelity for a given momentum may be obtained using
Eqs. \ref{hup}, \ref{hdown}.

In Fig. \ref{fig2} we show the $k$-space fidelity for two density matrices
that correspond to states in phases with different Chern numbers that
are separated by a transition such that the spectrum gap closes at the
point $\bs{k}=(\pi,0)$ (and equivalent points). As obtained before numerically,
the fidelity vanishes at these momenta values \cite{tharnier}.
Similar results may be obtained for other examples, as shown in Ref. \onlinecite{tharnier}.
We also consider in the left panel of Fig. \ref{fig3} two density matrices that correspond to a transition
from a trivial phase at $\mu_1=-6$ to another trivial phase at $\mu_2=6$ and $M_{z,1}=M_{z,2}=0.5$.
In the right panel of the same figure we consider a density matrix at a topological
phase with $C=-1$ to the same final trivial phase with $C=0$. Tracing a straigth line
between the initial and final points in the phase diagram we see that in the left
panel we cross twice gapless points at $\bs{k}=(0,0),(\pi,0),(\pi,\pi)$. In the
case of the right panel we cross gapless points at $\bs{k}=(0,0)$ (once), $\bs{k}=(\pi,\pi)$
(twice) and $\bs{k}=(\pi,0)$ (once).   

Taking the neighborhood of a point where the k-fidelity vanishes one can show
that there is a factor proportional to the momentum displacement from the
gapless point for each term that vanishes. Therefore evaluating the fidelity
between two points in the phase diagram such as, for instance, $M_z=0.5$ and $\mu_1=-2, \mu_2=2$
there is factor proportional to $k$ coming from each spin contribution leading
to a factor of $k^2$, and therefore a quadratic dispersion. In the case of Fig. \ref{fig2}
the dispersion is linear near $\bs{k}=(\pi,0)$ since the gapless point is only
crossed once and looking at Fig. \ref{fig3} we the spectrum is quadratic near all gapless
points in the left panel and linear near $\bs{k}=(0,0),(\pi,0)$ in the right panel and
quadratic near $\bs{k}=(\pi,\pi)$.
These results just confirm those obtained numerically before \cite{tharnier}.

This result is explained next.

\subsection{General result for $2 \times 2$ Hamiltonian matrix}
\label{subsec:general_result}

Given a set $\mathcal Q$ of Hamiltonian parameters (which, in case of, say, effective Hamiltonians, may include temperature as well), for each momentum $\bs{k}$ we have Hamiltonians $H_{q}(\bs{k})$ and the corresponding Gibbs states $\rho_{q}(\bs{k}) = e^{-\beta H_{q}(\bs{k})}/Z(\bs{k})$, with $q\in \mathcal Q$ and $\beta$ being the inverse temperature. Consequently, we are given the fidelity $F_{12}(\bs{k}) = F(\rho_{q_1}(\bs{k}), \rho_{q_2}(\bs{k}))$ and the energies $E_{q}(\bs{k})$.

At $\beta\rightarrow \infty$ limit, we are interested in finding the relation between the pairs of parameter points $(q_1,q_2)$ for which the fidelity vanishes, $F_{12}(\bs{k}) = 0$, and the existence of critical gapless points $q_c$, for which $E_{q_c}(\bs{k}) = 0$.

Given the Hamiltonian 
\be\label{eq:hamiltonian}
H_{q}(\bs{k})= \bs{h}_{q}(\bs{k}) \cdot \bs{\sigma},
\ee
its eigenvalues are $E_{q}^\pm (\bs{k}) = \pm |\bs{h}_{q}(\bs{k})|$, and the fidelity is 
\be 
F_{12}(\bs{k}) =  
\sqrt{ \frac{1}{2} \left( 1+ \frac{\bs{h}_{q_1}(\bs{k})}{|\bs{h}_{q_1}(\bs{k})|} \cdot
\frac{\bs{h}_{q_2}(\bs{k})}{|\bs{h}_{q_2}(\bs{k})|} \right) }. 
\label{central}
\ee
Thus, gapless points are given by 
\be \label{eq:gapless}
E_{q_c}(\bs{k}) = |\bs{h}_{q_c}(\bs{k})| = 0,
\ee
and the condition that the fidelity vanishes translates to
\be \label{eq:zero_fid}
\bs{h}_{q_1}(\bs{k}) \cdot \bs{h}_{q_2}(\bs{k}) = -|\bs{h}_{q_1}(\bs{k})| |\bs{h}_{q_2}(\bs{k})|,
\ee
which implies that the angle between $\bs{h}_{q_1}(\bs{k})$ and $\bs{h}_{q_2}(\bs{k})$
is $\pi$. 

This observation hints at the existence of the ``gapless vector'' $\bs{h}_{q_c}(\bs{k}) = \bs{0}$ between $\bs{h}_{q_1}(\bs{k})$ and $\bs{h}_{q_2}(\bs{k})$, defining the critical point $q_c$ of potential quantum phase transition (see section~\ref{sec:absence} for examples in which gapless excitations are not accompanied by transition lines). 
Nevertheless, a simple ``rotational'' counter example shows that, at least in principle, there exist models for which there exists no $q=q_c$ for which $\bs{h}_{q_c}(\bs{k}) = \bs{0}$. Indeed, consider $\bs{h}_{\varphi}(\bs{k}) = \cos(\varphi) \bs{e_x} + \sin(\varphi) \bs{e_y}$, for $q = \varphi \in [0,2\pi)$. We see that in this model there exist no gapless points, as the two energy bands are flat, $E_{q}^\pm (\bs{k}) = \pm 1$, while for each two ``antipodal'' parameter points $\varphi$ and $\varphi + \pi$, we have that the corresponding fidelity is zero.

A simple sufficient condition that allows to infer gapless points~\eqref{eq:gapless} from the existence of zero fidelity pairs~\eqref{eq:zero_fid} is the linearity, with respect to $q$, of the function $\bs{h}_{q}(\bs{k})$, {\em provided} that the set of parameters $\mathcal Q$ is not too restricted. Indeed, condition~\eqref{eq:zero_fid} implies $\bs{h}_{q_2}(\bs{k}) = -\lambda \bs{h}_{q_1}(\bs{k})$, for some positive $\lambda$. Assuming that $\mathcal Q$ is a subspace of a real linear space, define $q_c = \mu q_1 + \nu q_2$, for some $\mu,\nu \in \mathbb{R}$. Assuming linearity, we have
\begin{equation}
	\bs{h}_{q_c}(\bs{k}) = \bs{h}_{(\mu q_1 + \nu q_2)}(\bs{k}) = (\mu - \lambda\nu)\bs{h}_{q_1}(\bs{k}).
\end{equation}
To satisfy~\eqref{eq:gapless}, we need to satisfy
\begin{equation}
	\mu = \lambda\nu,
\end{equation}
which gives a line of critical points $q_c (\nu) = (\lambda q_1 +q_2 )\nu$, parametrized by $\nu$ (note the above disclaimer -- we require that for at least one $\nu\in \mathcal Q$, we also have $(\lambda q_1 +q_2 )\nu \in \mathcal Q$).

We show the above statement on the examples of topological insulator and $1d$ Kitaev model of spinless fermions (subsections~\ref{subsec:top_insulator} and~\ref{subsec:kitaev}, respectively), as well as for BCS superconductor, Ising model in a transverse field and graphene (Appendices~\ref{subsec:bcs}, \ref{subsec:ising} and \ref{subsec:graphene}, respectively). Nevertheless, linearity of $\bs{h}_{q}(\bs{k})$ with respect to $q$ is not the necessary condition for the existence of gapless points. In fact, the above ``rotational'' counter example can also hint towards a nonlinear model in which the condition~\eqref{eq:zero_fid} implies the existence of critical points $q_c$ satisfying~\eqref{eq:gapless}: in addition to $\varphi$, introduce parameter $\rho \in [0,+\infty)$, and define $\bs{h}_{(\rho, \varphi)} = \rho \cos(\varphi) \bs{e_x} + \rho\sin(\varphi) \bs{e_y}$. Note again that here, similarly to the case of linear dependance $\bs{h}_{q}(\bs{k})$ on $q$, to avoid the existence of gapless points, we had to restrict the parameters from $ q =(\rho , \varphi)$ to $q = \varphi$. As an example of concrete physical model, in subsection~\ref{subsec:haldane} we analyze below Haldane Chern insulator.

One could pose an ``opposite'' question, whether the existence of a gapless point $q_c$, for which $E_{q_c}(\bs{k}) = |\bs{h}_{q_c}(\bs{k})| = 0$, implies the existence of pairs of parameters $(q_1,q_2)$ for which the corresponding vectors $\bs{h}_{q_1}(\bs{k})$ and $\bs{h}_{q_2}(\bs{k})$  satisfy~\eqref{eq:zero_fid}, for which the fidelity vanishes, $F_{12}(\bs{k}) = 0$. A simple counterexample shows that this, in general, is not the case. Take $q=(\rho,\varphi)$, such that $\rho \in [0,1]$ and $\varphi \in [0,\pi/2]$. Define $\bs{h}_{(\rho,\varphi)} = \rho\cos(\varphi)\bs{e_x} + \rho\sin(\varphi)\bs{e_y}$. There is one gapless point, $\rho = 0$, but the fidelity is never zero, regardless of $(q_1,q_2)$. Again, as in the above counter-examples, the way to avoid the existence of ``zero-fidelity pairs'' $(q_1,q_2)$ is {\em to restrict}, this time the co-domain of the mapping $\bs{h}_{q}(\bs{k})$ to a set that excludes the existence of {\em any two} pairs of vectors for which $\bs{h}_{q_2}(\bs{k}) = -\lambda \bs{h}_{q_1}(\bs{k})$. Otherwise, having $\bs{h}_{q_c}(\bs{k}) = \bs{0}$, for some $q_c$, we can always find $q_1$ and $q_2$ for which
\bea
\bs{h}_{q_1}(\bs{k}) &=& \bs{h}_{q_c}(\bs{k}) + \delta \bs{h}_{q_1}(\bs{k}) \nonumber \\
\bs{h}_{q_2}(\bs{k}) &=& \bs{h}_{q_c}(\bs{k}) + \delta \bs{h}_{q_2}(\bs{k}).
\eea
Then, it follows that
\be
\bs{h}_{q_1}(\bs{k}) \cdot \bs{h}_{q_2}(\bs{k}) = \delta \bs{h}_{q_1}(\bs{k}) \cdot \delta \bs{h}_{q_2}(\bs{k}).
\ee

Consider for instance the Sato and Fujimoto model simplified to the
case where $\Delta_s=\alpha=0$, since the topological properties are not changed.
The transitions between the various phases occur at the momentum points
$\bs{k}=(0,0),(0,\pi),(\pi,\pi)$ (and equivalent points).
Consider for instance the point $\bs{k}=(0,0)$. The gapless point implies that
\be
4t+\mu+M_z=0
\ee
The vanishing of the fidelity implies that
\be
-1 = \text{sgn} (4t_1+\mu_1+M_{z,1} ) \text{sgn} (4t_2 +\mu_2 +M_{z,2} )
\ee
which is satisfied if the signals are opposite. The transitions at the
momentum origin occur in the vicinity of $\mu=-4$ if the magnetization is small
(we fix $t=1$). Similar expressions can be obtained in the vicinity of other
transition lines.

\section{Application to other systems}
\label{sec:application}

\begin{figure}
\includegraphics[width=0.45\columnwidth]{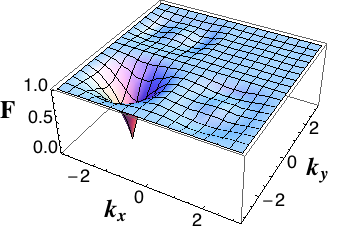}
\includegraphics[width=0.45\columnwidth]{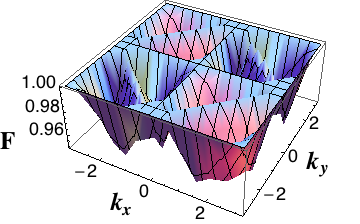}
\caption{\label{fig6}
(Color online) $k$-space fidelity for the topological insulator
of Eq. \ref{ti1}
with (left panel) $t_{x,1,1}=t_{x,1,2}=t_{y,1,1}=t_{y,1,2}=1,
t_{2,1}=1.2, t_{2,2}=0.3, t_{1,1}^{\prime}=t_{1,2}^{\prime}=0.5, 
\delta_1=\delta_2=0.1$. 
In the right panel we take $t_{x,1,1}=t_{x,1,2}=1, t_{y,1,1}=t_{y,1,2}=0,
t_{2,1}=1.2, t_{2,2}=0.3, t_{1,1}^{\prime}=t_{1,2}^{\prime}=0, 
\delta_1=\delta_2=0$. 
}
\end{figure}

\subsection{Topological insulator}
\label{subsec:top_insulator}

A simple toy model for a two-dimensional topological insulator with two bands
may be written as \cite{change}
\bea
h_x &=& \sqrt{2} t_{x,1} \left( \cos k_x+\cos k_y \right) \nonumber \\
h_y &=& \sqrt{2} t_{y,1} \left( \cos k_x-\cos k_y \right) \nonumber \\
h_z &=& 4 t_2 \sin k_x \sin k_y +2 t_i^{\prime} \left( \sin k_x +\sin k_y \right) +\delta
\label{ti1}
\eea
The terms $h_y$ and $t_1^{\prime}$ break time reversal symmetry and the $t_2$ term
breaks inversion symmetry. Since the system is two-dimensional, the system displays
regimes with non-vanishing Chern numbers. For instance, 
\bea
t_2 > t_1^{\prime}-\frac{\delta}{4} &,& C=2 \nonumber \\
t_2 < t_1^{\prime}-\frac{\delta}{4} &,& C=1 
\eea
At the points $\bs{k}=(\pm \pi/2,\pm \pi/2)$ both $h_x$ and $h_y$ vanish.
Around these points and taking $\delta=0$, $h_z$ has the form $h_z=4t_2-4t_1^{\prime}$ at $(-\pi/2,-\pi/2)$,
$h_z=4t_2+4t_1^{\prime}$ at $(\pi/2,\pi/2)$ and $h_z=-4t_2$ at the remaining points
$(\pi/2,-\pi/2), (-\pi/2,\pi/2)$. Therefore the momentum value that is associated
with the transition from $t_2>t_1^{\prime}$ to $t_2<t_1^{\prime}$ is the one where the gap
closes and the fidelity vanishes.

\begin{figure}
\includegraphics[width=0.75\columnwidth]{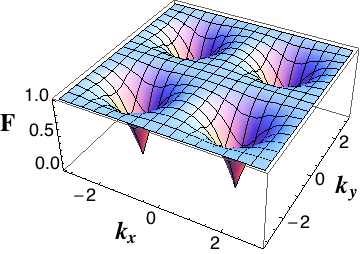}
\caption{\label{fig7}
(Color online) Fidelity for topological insulator of Eq. \ref{ti2}: $t_{2,1}=1, t_{2,2}=-1$.
}
\end{figure}

In Fig. \ref{fig6} we show the $k$-space fidelity for this toy model. 
In the left panel we consider two phases such that
phase 1 has $C=2$ and phase 2 has $C=1$, as discussed above.
In the right panel we consider an example  
where there is no time reversal symmetry breaking.
Also, there is no gapless point between the two sets of parameters.
Therefore the fidelity has no zeros.

Another simple toy model that involves a transition between two topological
regimes with $C=2$ and $C=-2$ is the Hamiltonian
\bea
h_x &=& \cos k_x + \cos k_y \nonumber \\
h_y &=& \cos k_x - \cos k_y \nonumber \\
h_z &=& t_2 \sin k_x \sin k_y 
\label{ti2}
\eea
The two regimes are obtained changing the sign of $t_2$.
The gap closes at the points $\bs{k}=(\pi/2,\pi/2),(\pi/2,-\pi/2)$, and equivalent points.
This is clearly shown by the fidelity in Fig. \ref{fig7}.

\subsection{$1d$ Kitaev model of spinless fermions}
\label{subsec:kitaev}

In momentum space we may write the Kitaev model \cite{kitaev} as
\bea
\hat H &=& \frac 1 2 \sum_k  \left( c_k^\dagger ,c_{-k}   \right)
\left(\begin{array}{cc}
\epsilon_k -\mu & -2 i \Delta \sin k \\
2 i \Delta \sin k & -\epsilon_k +\mu  \end{array}\right)
\left( \begin{array}{c}
c_{k} \\  c_{-k}^\dagger  \end{array}
\right) \nonumber \\
& &
\eea
with $\epsilon_k=-2t \cos k$. 
The Hamiltonian may be written using the Pauli matrices with
\be
\bs{h} =(0,2 \Delta \sin k,\epsilon_k-\mu)
\ee
The eigenvalues are therefore $\pm |\bs{h}|$, where
\be
|\bs{h}| = \sqrt{4\Delta^2\sin^2 k + (-2 t \cos k-\mu)^2}
\ee
The transitions lines occur for $\mu=2$ and $k=\pi$, for $\mu=-2$ and $k=0$ and
for $\Delta=0$ and $\cos k=-\mu/(2t)$. Therefore for $\Delta=0$ and $\mu=0$ the transition
occurs at $k=\pi/2$. It is easy to check the vanishing of the fidelity. For this
problem we can write that
\bea
\frac{\bs{h}_1 \cdot \bs{h}_2}{|\bs{h}_1| |\bs{h}_2|} &=&
\left( 4 \Delta_1 \Delta_2 \sin^2 k + (2t\cos k+\mu_1)(2t\cos k+\mu_2)\right)  \nonumber \\
&\times  & 1/\sqrt{4 \Delta_1^2 \sin^2 k + (2t\cos k+\mu_1)^2}
\nonumber \\
&\times & 1/\sqrt{4 \Delta_2^2 \sin^2 k + (2t\cos k+\mu_2)^2}
\eea
Considering for instance $\mu_1=\mu_2=0$ it is easilly seen that choosing for instance
$\Delta_1>0,\Delta_2<0$ the expression reduces to $-1$ (vanishing fidelity) if $k=\pi/2$,
which is the condition for the gapless point.

The vanishing of the fidelity may also be calculated directly using the eigenstates.
At the points $\mu=0, \Delta=\pm t$ the eigenvalues are $\pm 2$
and the eigenvectors are
\begin{eqnarray}
\psi_+ =
\text{sgn} \left[\cos \frac{k}{2}\right] \left(\begin{array}{c}
-i \frac{\Delta}{t} \sin \frac{k}{2} \\
\cos \frac{k}{2}   
   \end{array}\right),
\end{eqnarray}
\begin{eqnarray}
\psi_- =
\text{sgn} \left[\cos \frac{k}{2}\right] \left(\begin{array}{c}
\cos \frac{k}{2}   \\
-i \frac{\Delta}{t} \sin \frac{k}{2}
   \end{array}\right)
\end{eqnarray}
Taking now $\mu=0$ but any value of $\Delta$, the eigenvalues are
\be
\lambda_{\pm} = \pm 2 \sqrt{ (t \cos k)^2 +(\Delta \sin k)^2}
\ee
The eigenvectors are (see for example Ref. \onlinecite{aop})
\begin{eqnarray}
\psi_+^{\Delta} =
\left(\begin{array}{c}
\frac{-i 2\Delta \sin k}{\sqrt{2 \lambda_+ \left( \lambda_++2t \cos k \right)}}
   \\
\sqrt{\frac{\lambda_+ +2 t\cos k}{2 \lambda_+}} 
   \end{array}\right), 
\end{eqnarray}
\begin{eqnarray}
\psi_-^{\Delta} =
\left(\begin{array}{c}
\sqrt{\frac{\lambda_- -2 t\cos k}{2 \lambda_-}}
   \\
\frac{-i 2\Delta \sin k}{\sqrt{2 \lambda_- \left( \lambda_- -2t \cos k \right)}}
   \end{array}\right)
\end{eqnarray}
Consider for instance the states $\psi_+^{\Delta_1}$ and $\psi_+^{\Delta_2}$.
Consider $\Delta_1>0,\Delta_2<0$.
Their overlap is easily obtained
\bea
F_{12} &=& |-2\frac{|\Delta_1| |\Delta_2| \sin^2 k}{\sqrt{\lambda_{+,1} 
(\lambda_{+,1}+2t \cos k) \lambda_{+,2} (\lambda_{+,2}+2t\cos k)}}
\nonumber \\
&+& \sqrt{\frac{\lambda_{+,1}+2t\cos k}{2 \lambda_{+,1}}}
\sqrt{\frac{\lambda_{+,2}+2t\cos k}{2 \lambda_{+,2}}} |
\eea
At the momentum $k=\pi/2$ we get that $\lambda_+=2 |\Delta|$. Therefore
the fidelity vanishes.

\begin{figure}
\includegraphics[width=0.65\columnwidth]{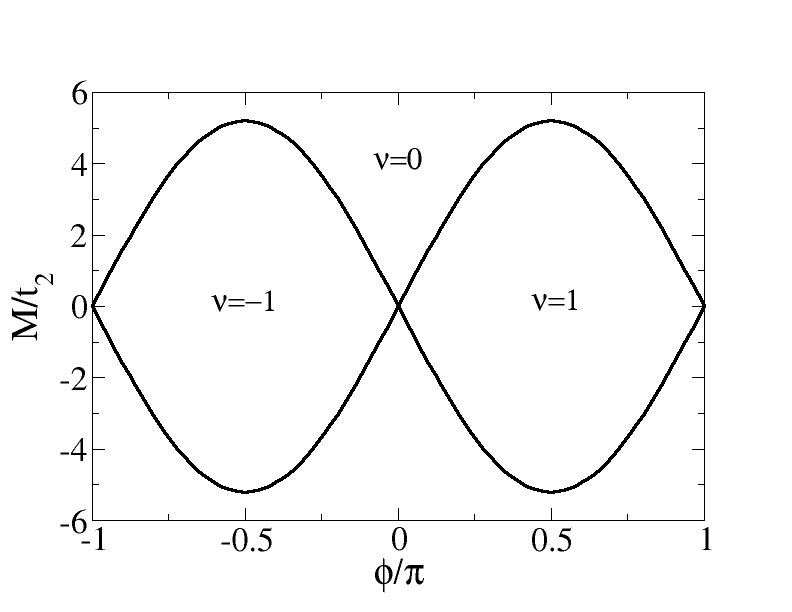}
\includegraphics[width=0.65\columnwidth]{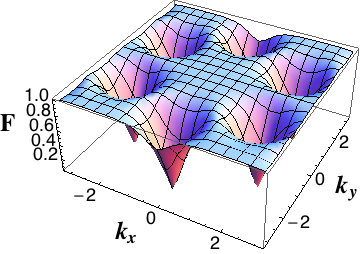}
\caption{\label{fig9}
(Color online) Fidelity for Haldane model with $t_{1}=1, t_{2}=0.25, m_1=m_2=0.25, \phi_1=\pi/2, \phi_2=-\pi/2$. 
}
\end{figure}

\subsection{Haldane Chern insulator}
\label{subsec:haldane}

We may also consider a graphene like model with the addition of hopping terms
between nearest-neighbors on the same sublattice, $t_2$, with a periodic magnetic flux
that breaks time-reversal inversal and therefore the possibility of a non-vanishing
Chern number (but with zero total flux through a unit cell). This generalization
was considered by Haldane \cite{haldane} as an example of a topological Chern
insulator in the absence of an external magnetic field. The magnetic flux is
included by adding a phase to the hopping amplitude $t_2$.
The Hamiltonian, including a mass term is given in momentum space by
\bea
H(\bs{k}) &=& 2 t_2 \cos \phi \sum_i \cos(\bs{k} \cdot \bs{b}_i ) I \nonumber \\
&+& t_1 \sum_i \left( \cos (\bs{k} \cdot \bs{a}_i ) \sigma_1 +
\sin (\bs{k} \cdot \bs{a}_i ) \sigma_2 \right) \nonumber \\
&+& \left( M -2t_2 \sin \phi \sum_i \sin (\bs{k} \cdot \bs{b}_i ) \right) \sigma_3
\eea
Here $t_1$ is the hopping between nearest-neighbors between one sublattice and the other,
$M$ is the mass term and the lattice vectors are $\bs{a}_1=(1,0), \bs{a}_2=(-1/2,\sqrt{3}/2),
\bs{a}_3=(-1/2,-\sqrt{3}/2)$ and $\bs{b}_1=\bs{a}_2-\bs{a}_3, \bs{b}_2=\bs{a}_3-\bs{a}_1,
\bs{b}_3=\bs{a}_1-\bs{a}_2$.
As shown in Fig. \ref{fig9}a, there are topological regions characterized by non-vanishing
Chern numbers $\pm 1$ for $|t_2/t_1|<1/3$ that lead to non-vanishing Hall conductances.
As a function of the phase $\phi$ and the gap magnitude, $M$, the non trivial phases occur
if $|M/t_2|<3\sqrt{3} |\sin \phi |$. 

The fidelity may be calculated diagonalizing the Hamiltonian and by direct evaluation of
the absolute value of the overlap of the eigenfucntions for two different sets of parameters.
As for the above models, the $k$-space fidelity vanishes when comparing two states
that are in two distinct phases that can be connected by a straight line in the phase diagram
that cuts a transition or transition lines. As an example, we show in Fig. \ref{fig9}b the
fidelity between states that differ by the value of $\phi=\pi/2,-\pi/2$ with the other
parameters fixed at $t_1=1,t_2=0.25,M=0.5$. The fidelity has zeros at the six corners of
the Brillouin zone. This occurs because as one crosses from $\nu=1$ to $\nu=0$ and
from $\nu=0$ to $\nu=-1$, each transition line is characterized either by the Dirac zeros
at $\bs{K}$ (and equivalent points) or $\bs{K}^{\prime}$ (and equivalent points). Therefore
the fidelity is linear around each of the vanishing points, as discussed above.

\section{Generalization to higher dimensional Hamiltonians}
\label{sec:high_dim}

\subsection{Fidelity}
\label{subsec:fid_high_dim}

Consider a Hamiltonian of the form
\begin{align*}
H=\sum_{\mu=1}^{d}h^{\mu}\gamma_{\mu},
\end{align*}
where $\gamma_{\mu}$, $\mu=1,...,d$ are Hermitian matrices corresponding to an irreducible representation of a Clifford algebra over the field of the complex numbers with $d$ generators with Euclidean signature,
\begin{align*}
\gamma_{\mu}\gamma_{\nu}+\gamma_{\nu}\gamma_{\mu}=2\delta_{\mu\nu} I_{2^{n}},
\end{align*}
where $I_{2^{n}}$ is the $2^n\times 2^n$ identity matrix and $n=\lfloor d/2 \rfloor$. From now on we will drop the index of the dimension of the vector space on the identity matrix. These matrices satisfy
\begin{align*}
\tr \big(\gamma_{\mu}\gamma_{\nu}\big)= 2^{n}\delta_{\mu\nu}. 
\end{align*}
We have $H^2=h^2 I$ and, therefore, the eigenvalues are $\pm ||\bs{h}||$, with $||\bs{h}||^2\equiv 
\sum h^{\mu}h^{\mu}$. Let us assume $h\neq 0$ and, w.l.o.g., that $||\bs{h}||=1$, i.e. 
$\bs{h}=(h^{\mu})$ determines an element of the $d-1$ dimensional sphere $S^{d-1}$. Notice that
\begin{align*}
P=\frac{1}{2}\big(I - H\big),
\end{align*}
commutes with $H$ and is a projector. Similarly, $Q=I -P=(1/2)(I+H)$, also commutes with $H$ and is a projector. Moreover,
\begin{align*}
H= Q-P=I-2P.
\end{align*}
So $P$ corresponds to the projector onto the $-1$ eigenvalue sector and $Q$ corresponds to the projector onto the $+1$ eigenvalue sector.\\
\indent Associated to $P$ we can build a density matrix 
\begin{align*}
\rho(\bs{h})= \frac{P}{\tr\big(P\big)}=\frac{P}{2^{n-1}}.
\end{align*}
The fidelity between two such density matrices, which we denote by $F(\bs{h}_1,\bs{h}_2)$, is given by
\bea
F(\bs{h}_1,\bs{h}_2)&=& \tr\Big(\sqrt{\sqrt{\rho(\bs{h}_1)}\rho(\bs{h}_2)\sqrt{\rho(\bs{h}_1)}}\Big)
\nonumber \\
&=& \frac{1}{2^{n-1}}\tr\Big(\sqrt{P_1P_2P_1}\Big),
\eea
where we wrote $P_i=(1/2)(1-\sum_{\mu}h^{\mu}_i\gamma_{\mu})\equiv (1/2)(1-H_i) $, $i=1,2$. 
Now, using $H_1H_2+H_2H_1=2\langle \bs{h}_1,\bs{h}_2\rangle I$, 
with $\langle \bs{h}_1,\bs{h}_2\rangle=\sum h_1^{\mu}h_2^{\mu}$,
\begin{align*}
P_1 P_2 P_1&= \frac{1}{8}\Big( 2I -2H_1 -H_2 +H_1 H_2 +H_2H_1 -H_1 H_2 H_1\Big)\\
&=\frac{1}{8}\Big(2I -2 H_1  +2\langle \bs{h}_1,\bs{h}_2\rangle (I-H_1)\Big)\\
&=\frac{1}{2}\big(1+\langle \bs{h}_1,\bs{h}_2\rangle \big)P_1.
\end{align*}
So that,
\begin{align*}
F(\bs{h}_1,\bs{h}_2)=\sqrt{\frac{1}{2}\big(1+\langle \bs{h}_1,\bs{h}_2\rangle \big)}.
\end{align*}
similarly to the result in Eq. \ref{central}.
Therefore if $\bs{h}_1$ and $\bs{h}_2$ form an angle of $\pi$ (for instance, if they are 
antipodal), the fidelity will vanish.

\subsection{Example: 3D topological insulator}
\label{subsec:example_high_dim}

\begin{figure}[h!]
\center
\includegraphics[scale=0.4]{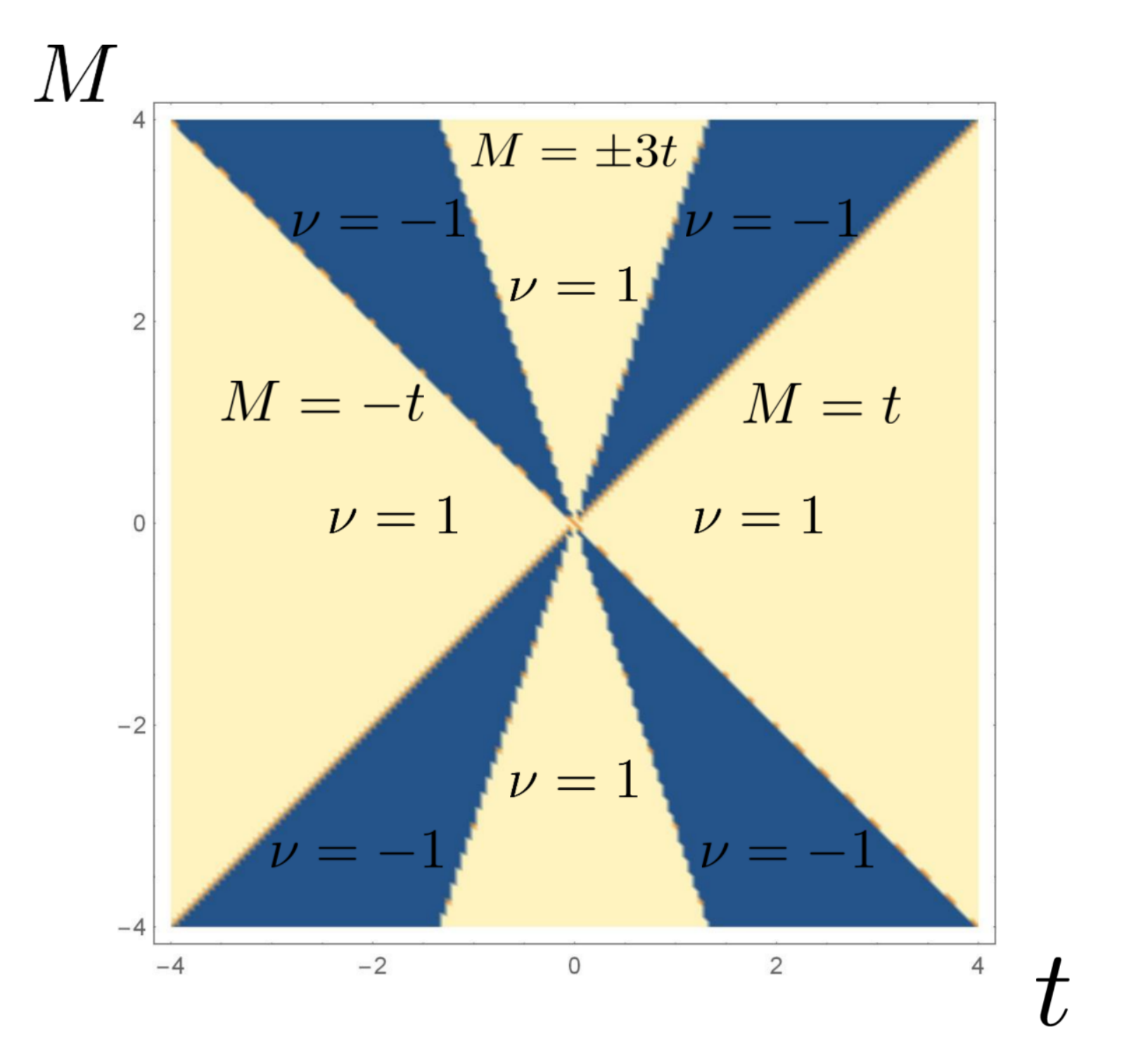}
\caption{The four lines $M=\pm t$ and $M=\pm 3t$ separate the phases where $\nu=\pm 1$.}
\label{fig:1}
\end{figure}

Consider the following model for a 3D topological insulator~\cite{pav:ryu:vis:10, mon:shi:11}
\begin{align*}
H(\bs{k})=v\tau^z\big(\sum_{\mu}\sigma^{\mu} \sin (k_\mu)\big)+\big(M-t\sum_{\mu}\cos(k_\mu)
\big)\tau^x,
\end{align*}
with $\mu=x,y,z$. The $\gamma$ matrices,
\begin{align*}
&\gamma_1=\tau^z\otimes \sigma^x=\left(\begin{array}{cc}
\sigma^x & 0\\
0 & -\sigma^x
\end{array}\right),\\
&\gamma_2=\tau^z\otimes \sigma^y=\left(\begin{array}{cc}
\sigma^y & 0\\
0 & -\sigma^y
\end{array}\right),\\
&\gamma_3=\tau^z\otimes \sigma^z=\left(\begin{array}{cc}
\sigma^z & 0\\
0 & -\sigma^z
\end{array}\right),\\
&\gamma_4=\tau^x\otimes I_2=\left(\begin{array}{cc}
0 & I_2\\
I_2 & 0
\end{array}\right),
\end{align*}
form an irreducible representation of a Clifford algebra in four generators with Euclidean signature. 
Our vector $\bs{h}$ is then given by
\begin{align*}
\bs{h}(\bs{k})=\big(v\sin(k_x),v\sin(k_y),v\sin(k_z),M-t\sum\cos(k_\mu)\big).
\end{align*}
The time-reversal operator is given by $\Theta=-i\big(I_2\otimes \sigma^y\big) K$, where $K$ is complex conjugation. Under time reversal $\bf{k}\to-\bf{k}$. The time-reversal invariant (TRI) momenta of the Brillouin zone $\text{B.Z.}\cong T^3$ are given by 
\begin{align*}
&(0, 0, 0),\ (\pi, 0, 0), \ (0 ,\pi ,0), \ (0 ,0 ,\pi),\\
&(\pi,\pi,0),\ (0,\pi,\pi),\ (\pi, 0 ,\pi), \ (\pi,\pi,\pi).
\end{align*}
The spatial inversion operator is given by $\Pi=\tau^x$. At a time-reversal invariant momentum $\bs{k}$, 
the Hamiltonian commutes with $\Pi$ and also
\begin{align*}
H(\bs{k})=(M-t\sum_\mu \cos(k_\mu))\Pi\equiv m(\bs{k})\Pi.
\end{align*}
The strong $\mathbb{Z}_2$ invariant is given by the product of the signs of the masses at TRI points,
\begin{align*}
\nu =\prod_{\{\bs{k}\in \text{B.Z.}:\bs{k}=-\bs{k}\}}\text{sgn}(m(\bs{k}))\in\mathbb{Z}_2. 
\end{align*}
Explicitly, it reads
\bea
\nu&=&\text{sgn}\Big[(M-3t)(M-t)^3(M+t)^3(M+3t)\Big]
\nonumber \\
&=&\text{sgn}\big[(M^2-9t^2)(M^2-t^2)\big].
\eea
The phase diagram is presented in Fig.~\ref{fig:1}.

\begin{figure*}
\includegraphics[width=0.32\textwidth]{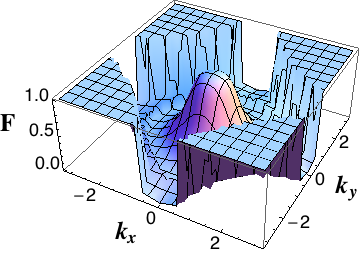}
\includegraphics[width=0.32\textwidth]{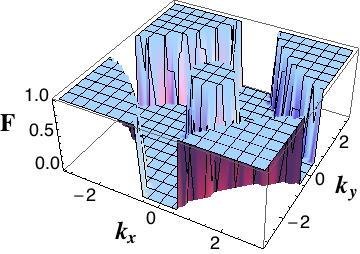}
\includegraphics[width=0.32\textwidth]{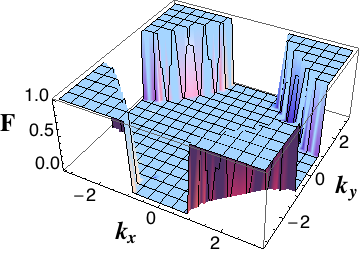}
\caption{\label{fig4}
(Color online) $k$-space fidelity for the $2d$ triplet superconductor with 
(left panel) $\Delta_{t,1}=0.6, \Delta_{t,2}=0., \mu_1=-3,\mu_2=-0.1, M_{z,1}=M_{z,2}=0.5, T=0$, 
(middle panel) $\Delta_{t,1}=0, \Delta_{t,2}=0., \mu_1=-3,\mu_2=-0.1, M_{z,1}=M_{z,2}=0.5, T=0$ and 
(right panel) $\Delta_{t,1}=0, \Delta_{t,2}=0., \mu_1=-3,\mu_2=-0.1, M_{z,1}=M_{z,2}=1.0, T=0$. 
}
\end{figure*}

We can now consider the $k$-space fidelity between groundstate subspaces associated 
with two Hamiltonians $H_1\equiv H(M_1,t_2)$ and $H_2\equiv H(M_2,t_2)$.
\begin{align*}
F(H_1(\bs{k}),H_2(\bs{k}))=\sqrt{\frac{1}{2}\Big(1+\frac{\langle \bs{h}_1(\bs{k}),\bs{h}_2(\bs{k})
\rangle}{||\bs{h}_1(\bs{k})||||\bs{h}_2(\bs{k})|| }\Big)}
\end{align*}
At TRI points we always have $H(\bs{k})=m(\bs{k})\Pi$, or, equivalently, 
$\bs{h}(\bs{k})=(0,0,0,m(\bs{k}))$. So for the fidelity to vanish at one point, 
we just need, for some TRI momenta $\bs{k}$
\begin{align*}
\text{sgn}(m_1(\bs{k})m_2(\bs{k}))=-1,
\end{align*}
i.e., the masses have opposite signs. In fact, this condition means that the 
elements of the three-dimensional $S^3$ defined by $\bs{h}_1(\bs{k})/|\bs{h}_1(\bs{k})|$ 
and $\bs{h}_2(\bs{k})/|\bs{h}_2(\bs{k})|$ are antipodal at this specific TRI momentum $\bs{k}$.
More strikingly, at these TRI momenta, the fidelity will always be either zero or one. 
A straight line connecting $\bs{h}_1$ and $\bs{h}_2$ with this property will always close 
the gap at this TRI momentum.\\
\indent For example, if we consider $v=1$, take $M_1=M_2=1$ and $t_1=1\pm s$, 
with $s=0.5$, we get at $\bs{k}=(0,0,\pi)$,
\begin{align*}
\frac{\langle \bs{h}_1(\bs{k}),\bs{h}_2(\bs{k})\rangle}{||\bs{h}_1(\bs{k})||||\bs{h}_2(\bs{k})|| }=-1.
\end{align*}
In fact this holds for any of the following TRI points:
\begin{align*}
&(0, 0, 0),\ (\pi, 0, 0), \ (0 ,\pi ,0), \ (0 ,0 ,\pi), \ (\pi,\pi,\pi).
\end{align*}
For $M=t$, the system is gapless at the TRI invariant momenta
\begin{align*}
(\pi,0,0),\ (0,\pi,0),\ (0,0,\pi).
\end{align*}
The fact that the masses have opposite signs in an odd number of points implies there was a topological phase transition in between, as confirmed by the phase diagram.

\section{Absence of transition lines and vanishing fidelity}
\label{sec:absence}

As  noted in subsection~\ref{subsec:general_result}, it is possible to find situations in which the fidelity vanishes, but this is not associated necessarily with a gapless point at some specific value of the coupling constants. 
Although we are primarily interested in the points of phase transitions, it is also 
interesting to analyze situations in which gapless points, in the presence of the vanishing fidelity, 
do not characterize some change of phase. We consider in the following two possible examples.

Consider first the $2d$ triplet superconductor studied above.
Turning off superconductivity, and since the normal term is not topological
the topological nature is destroyed. In Fig. \ref{fig4} we consider first (left panel)
a transition from the superconductor to a point where superconductivity is turned off.
The region where the fildelity vanishes widens but the overall features of
Fig. \ref{fig2} remain. However, turning off superconductivity alltogether, as shown
in the other panels of Fig. \ref{fig4}, the fidelity now vanishes in extended
zones that correspond to gapless points in regions of momentum space and that
are not associated with any transitions.

Another example is constructed as follows.
 Consider the tight binding of graphene. The Hamiltonian in momentum space is

\be
H(\bs{k})=t A(\bs{k})\sigma_{+} + t A^{*}(\bs{k})\sigma_{-} ,
\ee
where $t$ is a hopping amplitude, $\sigma$ are the pseudo-spin Pauli matrices and
\be
A(\bs{k})=\sum_{i=1}^{3} \exp(i\bs{k}\cdot \bs{a}_i),
\ee
where the $\bs{a}_i$ are nearest neighbour vectors. We can add a mass term, which amounts to taking
\be
H(\bs{k})\longrightarrow H(\bs{k}) +m\sigma_z.
\ee
Bring a parameter $\theta$, such that  $t=t(\theta)=t_0\cos(\theta)$ and $M=M(\theta)=t_0\sin(\theta)$. 
For $\theta=0$,
we have a vector $\bs{h}(\bs{k})=(t_0 \mbox{Re}A(\bs{k}),-t_0 \mbox{Im}A(\bs{k}),0)$ with gapless points at $\bs{K}$ and $\bs{K}^{\prime}$. 
For $\theta=\pi$, the vector
goes to $-h(\bs{k})$. Therefore, for $\bs{k}\neq \bs{K},\bs{K}^{\prime}$, we always have,
\be
F_{\bs{k}}(\theta_1=0,\theta_2=\pi)=0, 
\ee
for every $\bs{k}$. 
But by changing $\theta$, we will not have other gapless points other than $\bs{K},\bs{K}^{\prime}$,  
when $\theta=n\pi$, with $n$ integer,
since
\be
E(\bs{k})=\pm t_0 \sqrt{ \cos(\theta)^2 |A(\bs{k})|^2+\sin(\theta)^2}.
\ee

\section{Conclusions}
\label{sec:conclusions}

In this work we studied two-band models or more generally models that can be factorized to a set of two-bands. We investigated whether the $k$-space fidelity between states described by density matrices that correspond to points deep inside phases can provide information about the transition lines or sequence of transition lines that separate those phases. This extends previous numerical calculations for a $2d$ triplet spinful superconductor \cite{tharnier} where the result was identified. 

In particular, we analyzed the relation between the existence of vanishing points of the $k$-space fidelity and gapless points. We analyzed general $2 \times 2$ Hamiltonians and presented a sufficient condition for the existence of gapless points, given there are pairs of parameter points for which the fidelity between the corresponding states is zero. By presenting an explicit counter-example, we showed that the sufficient condition is not necessary. Further, we showed that, unless the set of parameter points is  suitably constrained, the existence of gapless points generically imply the accompanied pairs of parameter points with vanishing fidelity. 

We showed explicitly that the vanishing fidelity is accompanied by the gapless points of zero-temperature quantum phase transitions on a number of concrete models: a topological insulator, the $1d$ Kitaev model of spinless fermions, the BCS superconductor, the Ising model in a transverse field, graphene and the Haldane model for a Chern insulator. 

General Dirac-like Hamiltonians were also considered. We observed that the fidelity has the same form as in the two-band case. As a consequence, the same type of behavior is found, i.e., the $k$-space fidelity can vanish for points arbitrarily far from each other in parameter space, for momenta where the gap is found to close along a straight line joining the two points. As an example of this more general scenario, we considered a 3D topological insulator, classified by a $\mathbb{Z}_2$ topological invariant.

We also briefly discussed the finite-temperature case on the example of a $2d$ triplet superconductor. 

Finally, we presented examples of systems in which, although vanishing fidelity \emph{can} infer gapless points, those do not correspond to phase transition lines.

We conclude therefore that the results suggest that a vanishing fidelity strongly hints at a 
gapless point and eventually a transition between phases, but it does not hold in general and some
specific counter-examples can be found. 
Then one can do the established procedure of going through the phase diagram 
step by step to search for a singular point.

\section*{Acknowledgements}

The authors acknowledge discussions with 
Tharnier Puel de Oliveira, Pedro Ribeiro, V\'{\i}tor Rocha Vieira,
Nathan Goldman and Angelo Carollo.
Partial support from FCT through grant UID/CTM/04540/2013 
is acknowledged. 
BM thanks the support from Funda\c{c}\~ao para a Ci\^encia e Tecnologia
(Portugal) namely through programmes
PTDC/POPH/POCH and projects UID/EEA/50008/2013, IT/QuSim, IT/QuNet, ProQuNet, partially
funded by EU FEDER, from the EU FP7 project PAPETS (GA 323901) and from the JTF project
NQuN (ID 60478).
NP acknowledges the support of SQIG -- Security and Quantum Information Group, the Instituto de Telecomunica\c{c}\~oes (IT) Research Unit, ref. UID/EEA/50008/2013, the IT project QbigD funded by Funda\c{c}\~ao para a Ci\^encia e Tecnologia (FCT) PEst-OE/EEI/LA0008/2013, and the FCT project Confident PTDC/EEI-CTP/4503/2014. 


\appendix

\section{Zero-temperature applications to other systems}
\label{sec:other_applications}

\subsection{BCS superconductor}
\label{subsec:bcs}

Consider a conventional, non-topological, $s$-wave superconductor at finite temperature
described by the effective mean-field BCS Hamiltonian
\bea
\label{bcs-hamiltonian}
H_{BCS}^{eff} &=& \sum_{k} \varepsilon_k(n_{k\ua}+n_{-k\da}) 
\nonumber \\
&-& 
\sum_k(\Delta_k c_{k\ua}^\dag c_{-k\da}^\dag + \Delta^\ast_k 
c_{-k\da} c_{k\ua} - \Delta^\ast_k \langle c_{-k\da} c_{k\ua} \rangle ),
\nonumber \\
& &
\eea
To simplify we consider $\Delta_k=\Delta$ a parameter independent of momentum but an extended
$s$-wave superconductor could also be considered and it could also be determined
self-consistently. 
We will be interested in situations where $\rho_1$ and $\rho_2$ correspond
to points in parameter space, which we choose to be the temperature, $T$, and the
gap, $\Delta$, that are far apart and may be in the same or different
thermodynamic phases \cite{paunkovic2}.

\begin{figure}
\includegraphics[width=0.7\columnwidth]{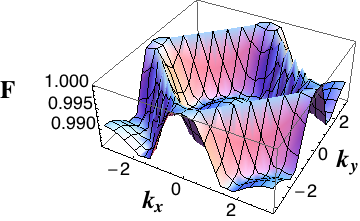}
\includegraphics[width=0.7\columnwidth]{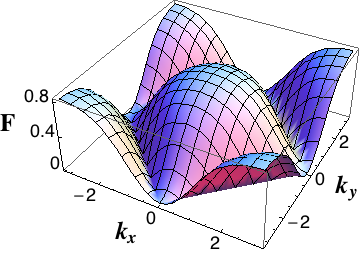}
\caption{\label{fig5}
(Color online) BCS superconductor: $\Delta_1=1,\Delta_2=0.5$ and $\Delta_1=1, \Delta_2=-1$.
}
\end{figure}

In Fig. \ref{fig5} 
we consider a transition between two points at $\mu=0$, one where the sign of
$\Delta$ does not change and one where $\Delta_1=-\Delta_2=1$. In the first case the
fidelity is close to one, as expected since we are in the same phase.
In the second case as $\Delta$ changes sign it crosses zero and there is a set
of gapless points and the fidelity vanishes at those points since
$\cos k_x+\cos k_y=0$.

\begin{figure}
\includegraphics[width=0.65\columnwidth]{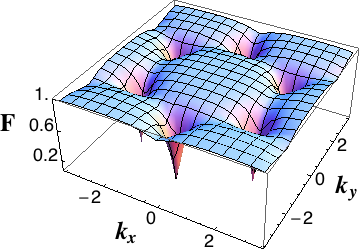}
\includegraphics[width=0.65\columnwidth]{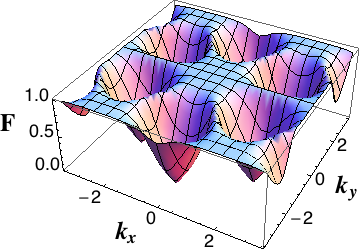}
\caption{\label{fig8}
(Color online) Fidelity for graphene with $m_1=-m_2=0.5$. In the top panel same mass on different
Dirac cones and lower panel opposite masses in different Dirac cones.
}
\end{figure}

\subsection{Ising model in a transverse field}
\label{subsec:ising}

The Ising model in a transverse field \cite{lieb,sen} described by
\be
H=-\sum_{j=1}^N \left( \sigma_j^x \sigma_{j+1}^x + h \sigma_j^z \right)
\ee
where $h$ is the transverse field,
can be related to the Kitaev model performing a Jordan-Wigner transformation \cite{jw}.
The fidelity between two states has been shown to be given by \cite{zanardi,gu}
\be
F(h,h^{\prime}) = \prod_{k\geq 0} \cos \left( \theta_k - \theta_k^{\prime} \right)
\ee
where the Bogoliubov angles are defined (for two values of the transverse field) in the form
\bea
\cos (2 \theta_k ) &=& \frac{\cos k -h}{\sqrt{1-2 h \cos k +h^2}}
\nonumber \\ 
\sin (2 \theta_k ) &=& \frac{\sin k}{\sqrt{1-2 h \cos k +h^2}}
\eea
The energy spectrum is given by
\be
\epsilon_k=\sqrt{1-2h \cos k +h^2}
\ee
Taking $h=1$ the spectrum becomes gapless at $k=0$ and if $h=-1$ at $k=\pi$.
Taking $k=0$ we get that
\be
\cos (2 \theta_0) = \text{sgn} (1-h)
\ee
Therefore
\bea
\cos \theta_0 &=& \sqrt{\frac{1}{2} \left( 1+\text{sgn} (1-h)\right)} 
\nonumber \\
\sin \theta_0 &=& \sqrt{\frac{1}{2} \left( 1-\text{sgn} (1-h)\right)} 
\eea
Therefore
\bea
\cos \left( \theta_0-\theta_0^{\prime} \right) &=&
\sqrt{\frac{1}{2} \left( 1+\text{sgn} (1-h)\right)}
\sqrt{\frac{1}{2} \left( 1+\text{sgn} (1-h^{\prime})\right)} \nonumber \\
&+& \sqrt{\frac{1}{2} \left( 1-\text{sgn} (1-h)\right)}
\sqrt{\frac{1}{2} \left( 1-\text{sgn} (1-h^{\prime})\right)} \nonumber \\
& &
\eea
As a consequence if we choose two points on the $h$ axis such that the $\text{sgn} 
(1-h)=\text{sgn} (1-h^{\prime})$
the fidelity at $k=0$ is one while if they are different the fidelity vanishes.

\subsection{Graphene}
\label{subsec:graphene}

\begin{figure*}
\includegraphics[width=0.24\textwidth]{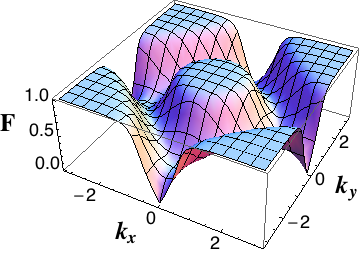}
\includegraphics[width=0.24\textwidth]{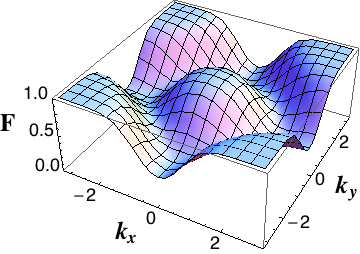}
\includegraphics[width=0.24\textwidth]{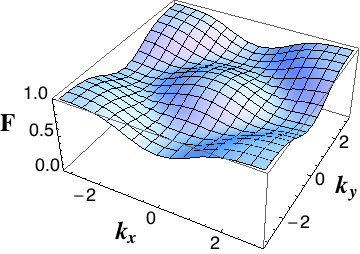}
\includegraphics[width=0.24\textwidth]{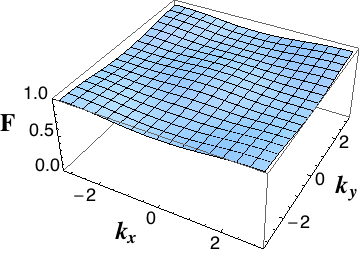}
\caption{\label{fig10} 
(Color online) 
Fidelity for the $2d$ triplet superconductor for 
$\Delta_{t,1}=\Delta_{t,2}=0.6, \mu_1=-2, \mu_2=0, M_{z,1}=M_{z,2}=0.5$ and several
temperatures: $T=0,0.25,0.5,1$.
}
\end{figure*}

We consider now a non-topological non-superconducting system such as graphene \cite{nunormp}. In order to gap
the spectrum we add a mass term.
We can model this massive graphene 
considering a mass term like $h_z=m$. In this case the system is non-topological
since, even though there is a non-trivial Berry curvature emerging from each Dirac cone, the total
Berry curvature cancels. However, introducing a mass term that depends on momentum such
that it has opposite signs at the two Dirac cones leads to a non-vanishing Berry curvature
and topological properties \cite{haldane}. We consider therefore in addition the case
\bea
h_x &=& 1+\cos \left(\sqrt{3} k_y \right) +\cos \left( \frac{\sqrt{3}}{2} k_y\right) 
\cos \left( \frac{3}{2} k_x \right) 
\nonumber \\
&-& \sin \left(\frac{\sqrt{3}}{2} k_y \right) \sin \left(\frac{3}{2} k_x
\right) 
\nonumber \\
h_y &=& \sin \left(\sqrt{3} k_y \right) +\sin \left( \frac{\sqrt{3}}{2} k_y\right) 
\cos \left( \frac{3}{2} k_x \right) \nonumber \\
&+& \cos \left(\frac{\sqrt{3}}{2} k_y \right) \sin \left(\frac{3}{2} k_x
\right) 
\nonumber \\
h_z &=& 4 m \sin \left(\frac{\sqrt{3}}{2}k_y \right) \left( \cos \left( \frac{3}{2} k_x \right)
-\cos \left( \frac{\sqrt{3}}{2} k_y \right) \right)
\nonumber \\
& & 
\eea
The Dirac points are situated at
$\bs{K} = \frac{2\pi}{3} \left(1,\frac{1}{\sqrt{3}} \right)$ and
$\bs{K}^{\prime} = \frac{2\pi}{3} \left(1,-\frac{1}{\sqrt{3}} \right)$
and $h_z(\bs{K})=-h_z(\bs{K}^{\prime})$.

In Fig. \ref{fig8} we consider the two cases where the mass is the same in both
Dirac cones or it changes sign. We consider $m_1=-m_2=0.5$.
Both models show vanishing fidelity
at the Dirac cones since by changing the sign of the mass at each Dirac point
implies a crossing through zero energy. 

\section{Temperature effects on $2d$ triplet superconductor}
\label{sec:temperature}

The effect of a finite temperature leads to a smoothning of the fidelity and the vanishing
points of the fidelity disappear. In Fig. \ref{fig10} we compare for the case of the
Sato and Fujimoto model the $k$-space fidelity for different temperatures. Even though
the vanishing points are absent, if the temperature is low there are signatures of
their locations, as expected.

\end{document}